# On "gauge renormalization" in classical electrodynamics

Alexander L. Kholmetskii
Belarusian State University
4, F. Skorina Avenue, Minsk 220080, Belarus
kholm@bsu.by

**Abstract**

In this paper we pay attention to the inconsistency in the derivation of the symmetric electromagnetic energy-momentum tensor for a system of charged particles from its canonical form, when the homogeneous Maxwell's equations are applied to the symmetrizing gauge transformation, while the non-homogeneous Maxwell's equations are used to obtain the motional equation. Applying the appropriate non-homogeneous Maxwell's equation to both operations, we have revealed an additional symmetric term in the tensor, named as "compensating term". Analyzing the structure of this "compensating term", we suggested a method of "gauge renormalization", which allows transforming the divergent terms of classical electrodynamics (infinite self-force, self-energy and self-momentum) to converging integrals. The motional equation obtained for a non-radiating charged particle does not contain its self-force, and the mass parameter includes the sum of mechanical and electromagnetic masses. The motional equation for a radiating particle also contains the sum of mechanical and electromagnetic masses, and does not yield any "runaway solutions". It has been shown that the energy flux in a free electromagnetic field is guided by the Poynting vector, whereas the energy flux in a bound EM field is described by the generalized Umov's vector, defined in the paper. The problem of "Poincaré stresses" is also examined. It has been shown that the presence of the "compensating term" in the electromagnetic energy-momentum tensor allows a solution of the "4/3 problem", where the total observable mass of the electron is completely determined by the Poincaré stresses and hence the conventional relativistic relationship between the energy and momentum is recovered.

## 1. Introduction

The problem of infinite electromagnetic (EM) mass of the electron and self-forces of charged particles has continued to be one of the central issues of classical electrodynamics during more than a century [1-8]. One of the reasons, explaining such a great attention to these problems, is their persistence in quantum electrodynamics [9, 10]. The simplest method to avoid the infinite EM mass of an electron is to add a compensating infinite negative mass. However, such a method does not overcome all difficulties of classical electrodynamics, in particular, the "runaway solutions" (*e.g.*, a "self-acceleration" of radiating electron). In addition, the total self-force of the electron includes its non-radiative and radiative parts, and the first of these is infinite. However, we cannot simply cancel the infinite self-action, because this inevitably would negate a radiative reaction observed experimentally. In the present paper we omit a detailed review of these problems, referring to the mentioned references [1-11], insofar as we will apply a primary modification of the energy-momentum tensor to remove an inconsistency, which seems not to have been revealed before.

It is known (see, *e.g.* [5, 6]) that the motional equation for an EM field with the Lagrangian density $-\frac{1}{16\pi}F_{\mu\nu}F^{\mu\nu}$ gives the following expression for the canonical energy-momentum tensor of the electromagnetic (EM) field

$$T_{EM}^{\mu\nu} = -\frac{1}{4\pi}\partial^\mu A^\gamma F^\nu{}_\gamma + \frac{1}{16\pi}g^{\mu\nu}F_{\gamma\alpha}F^{\gamma\alpha}, \qquad (1)$$

where $F^{\mu\nu} = \partial^\mu A^\nu - \partial^\nu A^\mu$ is the tensor of EM field, $A^\mu$ is the four-potential, $g^{\mu\nu}$ is the metric tensor, and $\mu, \nu = 0\ldots 3$. In order to transform Eq. (1) into a symmetric form, the gauge transformation

$$T_{EM}^{\mu\nu} \to T_{EM}^{\mu\nu} + \partial_\gamma \psi^{\mu\nu\gamma} \text{ (where } \psi^{\mu\nu\gamma} = -\psi^{\mu\gamma\nu}), \qquad (2)$$

should be applied. Choosing

$$\psi^{\mu\nu}{}_\gamma = \frac{1}{4\pi}A^\mu F^\nu{}_\gamma \qquad (3)$$

and writing

$$\partial^\gamma \psi^{\mu\nu}{}_\gamma = \frac{1}{4\pi}(\partial^\gamma A^\mu)F^\nu{}_\gamma + \frac{1}{4\pi}A^\mu(\partial^\gamma F^\nu{}_\gamma), \qquad (4)$$

we can transform the tensor (1) to the symmetric form

$$T_{EM}^{\mu\nu} = \frac{1}{4\pi}\left(-F^{\mu\gamma}F^\nu{}_\gamma + \frac{1}{4}g^{\mu\nu}F_{\gamma\alpha}F^{\gamma\alpha}\right), \qquad (5)$$

if we recognize that

$$\partial_\gamma F^{\nu\gamma} = 0 \qquad (6)$$

(the field equation in the absence of source charges). Eq. (5) represents the conventional expression for the tensor of EM field. Hereinafter we assume an empty space-time, where the metric tensor is Minkowskian.

Further, it is known that the energy-momentum tensor for matter has the form

$$T_M^{\mu\nu} = mc\frac{dx^\mu}{dt}\frac{dx^\nu}{d\tau}, \qquad (7)$$

where $m$ is the mass density, and $\tau$ is the proper time. Then the total energy-momentum tensor is defined as the sum of Eqs. (5) and (7):

$$T^{\mu\nu} = T_M^{\mu\nu} + T_{EM}^{\mu\nu} = mc\frac{dx^\mu}{dt}\frac{dx^\nu}{d\tau} + \left(-\frac{1}{4\pi}F^{\mu\gamma}F^\nu{}_\gamma + \frac{1}{16\pi}g^{\mu\nu}F_{\gamma\alpha}F^{\gamma\alpha}\right). \qquad (8)$$

The energy-momentum conservation law requires that the four-divergence of $T^{\mu\nu}$ should vanish:

$$\partial_\mu\left[(T_M)^\mu{}_\nu + (T_{EM})^\mu{}_\nu\right] = 0. \qquad (9)$$

Using the Maxwell' equations $\partial_\gamma F_{\mu\nu} = -\partial_\mu F_{\nu\gamma} - \partial_\nu F_{\gamma\mu}$, and

$$\partial_\gamma F^{\gamma\nu} = \frac{4\pi}{c}j^\nu \qquad (10)$$

($j^\nu$ is the four-current), we find

$$\partial_\mu (T_{EM})^\mu{}_\nu = -\frac{1}{c}F_{\nu\gamma}j^\gamma. \qquad (11)$$

Substituting Eq. (11) into Eq. (9), we derive the motional equation in the form

$$mc^2\frac{dv_\nu}{dt} = F_{\nu\gamma}j^\gamma, \qquad (12)$$

where $v_\nu$ is the four-velocity.

Eqs. (1)-(12) briefly reproduce the derivation of the energy-momentum tensor and motional equation from [5, 6], which are widely accepted. Then applying Eq. (12) to a single isolated charged particle we obtain the spatial components of this equation in the form

$$\frac{d\vec{p}}{dt} = q\vec{E} + q\frac{\vec{v}\times\vec{B}}{c}, \qquad (13)$$

where $q$ is the charge of particle, $\vec{p}$ is its momentum, $\vec{v}$ is the velocity, and $\vec{E}$, $\vec{B}$ are **its own** electric and magnetic fields. Furthermore, the requirement $\partial_\mu T^{\mu 0} = 0$ gives the following energy balance equation:

$$\frac{\partial u}{\partial t} + \nabla \cdot \vec{S} + \vec{E} \cdot \vec{j} = 0, \qquad (14)$$

where

$$u = \frac{E^2 + B^2}{8\pi} \qquad (15)$$

is the energy density of EM field of the particle, and

$$\vec{S} = \frac{c}{4\pi}\left(\vec{E} \times \vec{B}\right) \qquad (16)$$

is the Poynting vector. The term $\vec{E} \cdot \vec{j}$ in Eq. (14) describes **the self-action** of charged particle.

Usually the divergences of Eqs. (13) and (14) for a single isolated particle are related to the intrinsic inconsistency of classical EM theory, and they are overcome by the standard renormalization technique, mentioned above.

In this paper we intend to resolve the problems of self-action and infinite self-energy in classical electrodynamics, applying a procedure of renormalization of the energy-momentum tensor under its proper gauge transformation.

First of all we pay attention to a lack of logic in the derivation of Eq. (5) and further calculation of the four-divergence of $T^{\mu\nu}$. Namely, under the gauge transformation from Eq. (1) to Eq. (5) the homogeneous Maxwell equation (6) was used, while in proving the equality (11) the non-homogeneous Maxwell equation (10) was used. Thus, two different equations, (6) and (10), have been applied to the same physical entity, the electromagnetic energy-momentum tensor. Then it remains unclear whether this tensor describes free EM fields (as assumed in [5]), or is valid in the general case (as assumed in [6]). The revealed inconsistency prompts a closer look at the procedure of symmetrization of EM tensor, which is done in section 2. As a result of this analysis, a method of "gauge renormalization" is suggested. In section 3 we explore the motional equation and the energy-momentum conservation law in classical electrodynamics, obtained after the "gauge renormalization". In addition, a possible resolution of the problems of "4/3" and infinite self-energy of electron has been proposed. Finally, section 4 presents some conclusions.

## 2. The electromagnetic energy-momentum tensor for a system of charged particles and its "gauge renormalization"

Consider a system of $N>1$ charged particles, where the total tensor of the EM field $F^{\mu\nu}$ represents the sum of corresponding tensors $f^{\mu\nu}_{(k)}$ for each particle

$$F^{\mu\nu} = \sum_k f^{\mu\nu}_{(k)} \qquad (17)$$

($k=1\ldots N$) due to the superposition principle. The mechanical energy-momentum tensor (7) is properly modified as

$$T^{\mu\nu}_M = \sum_k m_{(k)} c \frac{d x^\mu_{(k)}}{dt} \frac{d x^\nu_{(k)}}{d\tau}, \qquad (18)$$

where the mass density is defined by the equation $m_{(k)} = M_{(k)} \delta(\vec{r} - \vec{r}_k)$, $M_{(k)}$ being the mass of particle $k$.

Determining the electromagnetic energy-momentum tensor for this system, we again proceed from the canonical form (1) and apply the gauge transformation (2). We use the gauge function (3) modified for the discrete system of $N$ particles:

$$\psi^{\mu\nu\gamma} = \frac{1}{4\pi} \sum_k A^{\mu}_{(k)} \sum_l f^{\nu\gamma}_{(l)} \qquad (19)$$

($l=1…N$). Noting that

$$\partial^{\gamma}\psi^{\mu\nu}{}_{\gamma} = \frac{1}{4\pi}\left(\sum_k \partial^{\gamma} A^{\mu}_{(k)}\right)\left(\sum_l f^{\nu}_{(l)\gamma}\right) + \frac{1}{4\pi}\left(\sum_k A^{\mu}_{(k)}\right)\left(\sum_l \partial^{\gamma} f^{\nu}_{(l)\gamma}\right),$$

and carrying out the gauge transformation (2) for the tensor (1), we obtain with account of Eq. (17):

$$T_{(EMG)}{}^{\mu\nu} = \frac{1}{4\pi}\left(-\sum_k f^{\mu\gamma}_{(k)} \sum_l f^{\nu}_{(l)\gamma} + \frac{1}{4}g^{\mu\nu}\sum_k f_{(k)\gamma\alpha}\sum_l f^{\gamma\alpha}_{(l)}\right) + \frac{1}{4\pi}\sum_k A^{\mu}_{(k)}\sum_l \partial^{\gamma} f^{\nu}_{(l)\gamma}. \qquad (20)$$

Eq. (20) differs from Eq. (5) by the term $\frac{1}{4\pi}\sum_k A^{\mu}_{(k)}\sum_l \partial^{\gamma} f^{\nu}_{(l)\gamma}$, which was omitted in Eq. (5) due to the condition (6), which cannot be accepted for the system of charged particles. In order to distinguish the tensor (20) from the conventional tensor (5), we name it as the "generalized electromagnetic energy-momentum tensor" (EMG).

Now let us apply the non-homogeneous Maxwell's equation (10) for any particle $l$. Outside of this particle $\partial^{\gamma} f^{\nu}_{(l)\gamma} = 0$, while at its location $\partial^{\gamma} f^{\nu}_{(l)\gamma} = -(4\pi/c)j^{\nu}_l$. Hence at this point the last term in *rhs* of Eq. (20) is equal to

$$(l)^{\mu\nu} = -\frac{1}{c}\left(\sum_k A^{\mu}_{(k)}\right) j^{\nu}_{(l)} = -\frac{1}{c} A^{\mu}_{(l)} j^{\nu}_{(l)}, \qquad (21)$$

insofar as the four-potential $A^{\mu}_{(l)}$ dominates over all other $A^{\mu}_{(k)}$ at the location of $l^{\text{th}}$ particle. Note that the tensor (21) is symmetrical, because $A^{\mu}_{(l)}$ is proportional to $v^{\mu}_{(l)}$, and its trace coincides with the Lagrangian of charged particle in an EM field, where an external field is replaced by its own EM field. We name the tensor (21) as **"compensating term"** for the reasons clarified below. Defining the same compensating term for each particle from the considered ensemble, we write the generalized electromagnetic energy-momentum tensor in the form:

$$T_{EMG}{}^{\mu\nu} = T_{EM}{}^{\mu\nu} - \frac{1}{c}\sum_k A^{\mu}_{(k)} j^{\nu}_{(k)}, \qquad (22)$$

where we denoted

$$T_{EM}{}^{\mu\nu} = \frac{1}{4\pi}\left(-\sum_k f^{\mu\gamma}_{(k)}\sum_l f^{\nu}_{(l)\gamma} + \frac{1}{4}g^{\mu\nu}\sum_k f_{(k)\gamma\alpha}\sum_l f^{\gamma\alpha}_{(l)}\right).$$

The latter represents the conventional electromagnetic energy-momentum tensor (5), modified for the discrete system of $N$ charged particles. We can rewrite this tensor in the form

$$T_{EM}{}^{\mu\nu} = T_{(EM)\text{ex}}{}^{\mu\nu} + \frac{1}{4\pi}\sum_k\left(-f^{\mu\gamma}_{(k)} f^{\nu}_{(k)\gamma} + \frac{1}{4}g^{\mu\nu} f_{(k)\gamma\alpha} f^{\gamma\alpha}_{(k)}\right),$$

where the tensor $T_{(EM)\text{ex}}{}^{\mu\nu}$ is defined by the equation

$$T_{(EM)\text{ex}}{}^{\mu\nu} = \frac{1}{4\pi}\left(-\sum_k f^{\mu\gamma}_{(k)}\sum_{l\neq k} f^{\nu}_{(l)\gamma} + \frac{1}{4}g^{\mu\nu}\sum_k f_{(k)\gamma\alpha}\sum_{l\neq k} f^{\gamma\alpha}_{(l)}\right). \qquad (23)$$

The introduced subscript "ex" indicates that the terms of "self-action", containing $(f_k)(f_k)$ ($k=1…N$), have been excluded from the tensor $T_{(EM)\text{ex}}{}^{\mu\nu}$. One can see that at the location of any particle $l$, this tensor satisfies the equality

$$\partial_{\mu}\left(T_{(EM)\text{ex}}\right)^{\mu}{}_{\nu} = -\frac{1}{c}\left(F_{\nu\gamma}\right)_{\text{ex}(l)} j^{\gamma}_{(l)}, \qquad (24)$$

where $(F_{\nu\gamma})_{ex(l)}$ does not contain $\left(f_{(l)}\right)_{\nu\gamma}$. Then Eq. (22) acquires the form

$$T_{EMG}{}^{\mu\nu} = T_{(EM)ex}{}^{\mu\nu} + \sum_k T_{(k)EEM}{}^{\mu\nu}, \qquad (25)$$

In the latter equation we have introduced a new tensor

$$T_{(k)EEM}{}^{\mu\nu} = \frac{1}{4\pi}\left(-f_{(k)}^{\mu\gamma} f_{(k)}^{\nu}{}_{\gamma} + \frac{1}{4} g^{\mu\nu} f_{(k)\gamma\alpha} f_{(k)}^{\gamma\alpha}\right) - \frac{1}{c} A_{(k)}^{\mu} j_{(k)}^{\nu}, \qquad (26)$$

which describes only the properties of particle $k$, but not its interaction with other particles. That is why we can name it as the **Eigen ElectroMagnetic (EEM) energy-momentum tensor** of charged particle, supplying it by the subscript "EEM".

Eqs. (22) and (25) can be derived in another way, using the energy-momentum tensor, defined according to Hilbert [6]:

$$\frac{1}{2}\sqrt{-g}\,T_{\mu\nu} = \frac{\partial \sqrt{-g}\,L}{\partial g^{\mu\nu}} - \frac{\partial}{\partial x^{\gamma}}\frac{\partial \sqrt{-g}\,L}{\partial \left(g^{\mu\nu}/\partial x^{\gamma}\right)}, \qquad (27)$$

where $L$ is the electromagnetic Lagrangian density. Taking $L$ in the form

$$L = -\frac{1}{c}\sum_k A_{(k)\mu} \sum_l j_{(l)}^{\mu} - \frac{1}{16\pi}\sum_k f_{(k)\gamma\alpha} \sum_l f_{(l)}^{\gamma\alpha}, \qquad (28)$$

with inclusion of both "interaction part" (the first term in *rhs* of Eq. (28)) and "field part" (the second term in *rhs* of Eq. (28)), and inserting $L$ from Eq. (28) into Eq. (27), we obtain the generalized electromagnetic energy-momentum tensor in the form

$$T_{(EMG)}{}^{\mu\nu} = \frac{1}{4\pi}\left(-\sum_k f_{(k)}^{\mu\gamma} \sum_l f_{(l)}^{\nu}{}_{\gamma} + \frac{1}{4} g^{\mu\nu} \sum_k f_{(k)\gamma\alpha} \sum_l f_{(l)}^{\gamma\alpha}\right) - \frac{1}{c}\sum_k A_{(k)}^{\mu} \sum_l j_{(l)}^{\nu}. \qquad (29)$$

(Under manipulation with Eqs. (27) and (28) we have used the equality [6] $\frac{\partial \sqrt{-g}}{\partial g_{\mu\nu}} = \frac{1}{2}\sqrt{-g}\,g^{\mu\nu}$).

We again see that outside the particles ($j_{(l)} = 0$) the second term in *rhs* of Eq. (29) vanishes, while at the location of each $l^{\text{th}}$ particle, $A_{(l)}$ dominates over the four-potentials of all other particles. Hence

$$\sum_k A_{(k)}^{\mu} \sum_l j_{(l)}^{\nu} = \sum_l A_{(l)}^{\mu} j_{(l)}^{\nu},$$

and Eq. (29) agrees with Eqs. (22) and (25).

Using the tensor (25) and taking into account the matter tensor (18), we write the total energy-momentum tensor as

$$T^{\mu\nu} = \left(\sum_{k=1}^{N} m_{(k)} c \frac{d x_{(k)}^{\mu}}{dt}\frac{d x_{(k)}^{\nu}}{d\tau} + T_{(k)EEM}{}^{\mu\nu}\right) + T_{(EM)ex}{}^{\mu\nu}. \qquad (30)$$

The above-introduced EEM tensor (26) represents the difference of two divergent terms and, in fact, is uncertain. Nevertheless, we can examine its general properties, considering first an isolated charged particle, moving at the constant velocity $\vec{v}$ in the frame of observation. For such a particle $T_{(EM)ex}{}^{\mu\nu} = 0$ by definition, and its total energy-momentum tensor acquires the form

$$T^{\mu\nu} = mc\frac{dx^{\mu}}{dt}\frac{dx^{\nu}}{d\tau} + T_{EEM}{}^{\mu\nu}, \qquad (31)$$

where its rest mechanical mass density is denoted as $m$. For the total energy-momentum tensor $\partial_{\mu} T^{\mu\nu} = 0$. Since for a freely moving particle $\partial_{\nu}\left(mc\frac{dx^{\mu}}{dt}\frac{dx^{\nu}}{d\tau}\right) = 0$, then

$$\partial_\mu T_{EEM}{}^{\mu\nu} = 0, \tag{32}$$

too. Hence we get the energy balance equation for a bound EM field of an isolated charged particle:

$$\frac{\partial T_{EEM}{}^{0\nu}}{\partial x^\nu} = \frac{1}{4\pi}\frac{\partial}{\partial x^\nu}\left[-(f_s)^{0\gamma}(f_s)^\nu{}_\gamma + \frac{1}{4}g^{0\nu}(f_s)_{\gamma\alpha}(f_s)^{\gamma\alpha}\right] - \frac{1}{c}\frac{\partial}{\partial x^\nu}\left[(A_s)^0(j_s)^\nu\right] = 0.$$

where the subscript "s" refers to an isolated charged particle. Further, using the equalities

$$\frac{1}{4\pi}\left[-(f_s)^{0\gamma}(f_s)^0{}_\gamma + \frac{1}{4}g^{00}(f_s)_{\gamma\alpha}(f_s)^{\gamma\alpha}\right] = \frac{E_s^2 + B_s^2}{8\pi} = u_s,$$

$$\frac{1}{4\pi}\left[-(f_s)^{i\gamma}(f_s)^0{}_\gamma + \frac{1}{4}g^{i0}(f_s)_{\gamma\alpha}(f_s)^{\gamma\alpha}\right] = \frac{c}{4\pi}(\vec{E}_s \times \vec{B}_s)^i = S_s^i \ (i=1\ldots 3),$$

we derive

$$\frac{\partial u_s}{\partial t} + \nabla \cdot \vec{S}_s - \frac{\partial}{\partial x^\mu}(A_s)^\mu \rho = 0. \tag{33}$$

where $\rho$ is the charge density of the particle. Using the vector identity $(\vec{E}_s \times \vec{B}_s) = \vec{B}_s \cdot (\nabla \times \vec{E}_s) - \vec{E}_s \cdot (\nabla \times \vec{B}_s)$ as well as the Maxwell's equations $(\nabla \times \vec{E}_s) = -\partial \vec{B}_s/c\partial t$, $(\nabla \times \vec{B}_s) = (4\pi/c)\vec{j}_s + \partial \vec{E}_s/c\partial t$, we find that

$$\nabla \cdot \vec{S}_s = -\frac{\partial u_s}{\partial t} - \vec{j}_s \cdot \vec{E}_s. \tag{34}$$

Combining Eqs. (33) and (34), one obtains:

$$-\vec{j}_s \cdot \vec{E}_s - \frac{\partial}{c\partial x^\nu}\left[(A_s)^0(j_s)^\nu\right] = 0 \text{ or}$$

$$\vec{j}_s \cdot \vec{E}_s + \frac{d}{dt}(\rho_s \varphi_s) = 0. \tag{35}$$

Outside the charged particle its charge density is identically equal to zero, and both terms in *lhs* of Eq. (35) disappear. Thus, the equality $\partial T_{EEM}{}^{0\nu}/\partial x^\nu = 0$ is valid in the whole free space. However, at the location of the particle the terms of Eq. (35) trend to infinity. Their vanishing sum signifies that the "self-work" done $\vec{j}_s \cdot \vec{E}_s$ is compensated by the change of the "potential energy" of particle $U_{ps}=\rho_s\varphi_s$. Noting that $\vec{j}_s \cdot \vec{E}_s = dE_{ks}/dt$, $E_{ks}$ being the kinetic energy, we arrive at the equality

$$\frac{d}{dt}(E_{ks} + U_{ps}) = 0,$$

which means the conservation of the sum of kinetic and potential energy. For an isolated charged particle both components of energy do not depend on time, and the particle moves at a constant velocity, as it should be. We underline that without the introduced "compensating term" (21) in the EEM tensor, we would obtain

$$\frac{\partial T_{EEM}{}^{0\nu}}{\partial x^\nu} = \vec{j}_s \cdot \vec{E}_s, \tag{36}$$

and the implementation of conservation law (32) would be impossible, if only the artificial requirement to equate to zero the divergent term of self-action $\vec{j}_s \cdot \vec{E}_s$ were applied.

In a similar way we can analyse the spatial components of Eq. (32). Outside the charged particle we get $\partial_\mu T_s{}^{\mu i} = 0$ for $i=1\ldots 3$. At the location of the particle

$$\frac{d}{dt}(\vec{S}_s - \rho_s \vec{A}_s) = 0,$$

which means that the time rate of the divergent Poynting vector $\vec{S}_s$ is compensated by the corresponding time rate of the divergent "potential momentum" $\rho_s \vec{A}_s$ of the particle. We again emphasize that without the compensating term (21) in the EEM tensor we would get

$$\frac{\partial T_{EEM}{}^{iv}}{\partial x^{\nu}} = \frac{d}{dt} S^{i}, \tag{37}$$

and the implementation of the conservation law (32) would be again impossible, if only the artificial requirement to equate to zero the divergent term $S^{i}$ were applied.

Nevertheless, cancelling a self-action for an isolated charged particle with the help of EEM tensor (26), we have still failed to determine unambiguously the total energy and momentum of such a particle. Indeed, Eq. (26) yields the following energy density of EM field at the location of the particle

$$T_{EEM}{}^{00} = \frac{E_s{}^2 + B_s{}^2}{8\pi} - \rho_s \varphi_s, \tag{38}$$

as well as the momentum density

$$\vec{p}_{EMs} = \frac{c}{4\pi}\left(\vec{E}_s \times \vec{B}_s\right) - \rho_s \vec{A}_s. \tag{39}$$

A vagueness of these quantities means the impossibility of determining the total energy and momentum of the EM field of a single particle.

Under these conditions we can carry out a suitable gauge modification of the EEM tensor (26), in order to escape the mentioned shortcomings. This mathematical problem can be much more easily solved physically, if we introduce a new tensor satisfying the conservation law (32). Namely, let us use a natural assumption that the total mass of a charged particle $M_t$ is composed from its mechanical mass $M$ and the mass $M_{EM}$ of its EM field. Denoting as $m$ and $m_{EM}$ the corresponding rest mass densities, we transform the matter tensor (7) to the form

$$T_M{}^{\mu\nu} = (m + m_{EM})c\frac{dx^{\mu}}{dt}\frac{dx^{\nu}}{d\tau}, \tag{40}$$

where for an isolated charged particle $\partial_{\mu}T_M{}^{\mu\nu} = 0$. Owing to the law of charge conservation, the mechanical mass cannot be transformed into EM mass and vice versa. Therefore, the vanished four-divergence is derived independently for the mechanical and EM parts of the tensor (40), and

$$\partial_{\mu}\left(m_{EM}c\frac{dx^{\mu}}{dt}\frac{dx^{\nu}}{d\tau}\right) = 0. \tag{41}$$

We see that the symmetric tensor

$$T_{mass}{}^{\mu\nu} = m_{EM}c\frac{dx^{\mu}}{dt}\frac{dx^{\nu}}{d\tau}, \tag{42}$$

named by us as **the tensor of EM mass**, also satisfies Eq. (32). Hence it is connected with the tensors $T_{(EEM)}{}^{\mu\nu}$ by the gauge transformation (2). Therefore, we can replace $T_{EEM}{}^{\mu\nu}$ by $T_{mass}{}^{\mu\nu}$ in equation (30) for the total energy-momentum tensor of the system of charged particles[1]:

---

[1] We underline that this gauge operation would be impossible, if the compensating term (21) were not added to the EEM tensor (26). Indeed, without the term (21) the trace of the EEM tensor (26) would be equal to zero, while the trace of tensor of EM mass (42) is not vanishing. Hence in no way can the two tensors $T_{EEM}{}^{\mu\nu}$ and $T_{mass}{}^{\mu\nu}$ be connected by the gauge transformation (2). One can also see that without the compensating term (21) Eq. (32) is reduced to Eqs. (36) and (37), which are not zero. The artificial requirement to exclude the terms of self-action and to equate Eq. (32) to zero does not influence its mathematical structure, giving $\partial_{\nu}\left[-(f_s)^{\mu\gamma}(f_s)^{\nu}{}_{\gamma} + \frac{1}{4}g^{\mu\nu}(f_s)_{\gamma\alpha}(f_s)^{\gamma\alpha}\right] \neq 0$. Therefore, we again conclude that the conventional tensor $\frac{1}{4\pi}\left[-(f_s)^{\mu\gamma}(f_s)^{\nu}{}_{\gamma} + \frac{1}{4}g^{\mu\nu}(f_s)_{\gamma\alpha}(f_s)^{\gamma\alpha}\right]$ and the tensor (42) of EM mass $T_{mass}{}^{\mu\nu}$ are not related by the gauge transformation (2), unless we add the compensating term (21) to the first tensor, and get the EEM tensor $T_{EEM}{}^{\mu\nu}$ in the adopted definition (26).

$$T^{\mu\nu} = \left(\sum_{k=1}^{N} c\left[m_{(k)} + m_{(k)EM}\right] \frac{d x_{(k)}^{\mu}}{dt} \frac{d x_{(k)}^{\nu}}{d\tau}\right) + T_{(EM)ex}^{\mu\nu}. \tag{43}$$

It is known that the gauge transformation (2) does not influence the total energy and momentum, and

$$\int_V T_{mass}^{00} dV = \int_V T_{EEM}^{00} dV \, ; \quad \int_V T_{mass}^{0i} dV = \int_V T_{EEM}^{0i} dV \tag{44), (45}$$

(the integration is carried out over the whole 3-space $V$). These equalities allow us to establish a relationship between the introduced EM mass of particle and its electric and magnetic fields. In particular, combining Eqs. (38), (42), and (44), we get for the rest frame of the charged particle:

$$M_{EM} = \frac{1}{c^2} \int_V \frac{E_s^2}{8\pi} dV - \frac{1}{c^2} \int_V \rho_s \varphi_s dV, \tag{46}$$

while combining Eqs. (39), (42), (45) we arrive at

$$\frac{M_{EM} \vec{v}}{\sqrt{1-v^2/c^2}} = \int_V \frac{1}{4\pi c}\left(\vec{E}_s \times \vec{B}_s\right) dV - \frac{\vec{v}}{\sqrt{1-v^2/c^2}} \int_V \frac{\rho_s \varphi_s}{c^2} dV. \tag{47}$$

These equations state that the difference of two divergent integrals in their *rhs* must be finite and equal to the EM mass of particle (Eq. (46)) and EM momentum of particle (Eq. (47)). Such statements are sufficient for further development of classical theory. Nevertheless, it seems interesting to extend a classical analysis of Eqs. (46), (47) to $r\to 0$. This will be done below in subsection 3.5 in relation to the "Poincaré stresses".

Thus, the obtained tensor (43) contains single-valued quantities and does not include a self-action of charged particles due to Eq. (24). The method proposed in this section can be termed a "gauge renormalization". We have to emphasize that this method has been applied to a bound EM field of a non-radiating isolated charged particle. If a particle moves in the external EM fields, and its EM radiation is not negligible, we have to proceed from the general tensor (25) for description of its EM field. Then the total energy-momentum tensor acquires the form

$$T^{\mu\nu} = \left(\sum_{k=1}^{N} c\, m_{(k)} \frac{d x_{(k)}^{\mu}}{dt} \frac{d x_{(k)}^{\nu}}{d\tau} + T_{(k)\,EEM}^{r\,\mu\nu}\right) + T_{(EM)ex}^{\mu\nu}, \tag{48}$$

where the superscript "r" indicates that the eigen electromagnetic energy-momentum tensor includes the radiation of each particle $k$. In order to write this EEM tensor explicitly, we use the superposition principle, whence the electromagnetic energy-momentum tensor of each particle represents the sum of components with a bound $f_b$ and free $f_r$ EM fields, with

$$\partial_\mu (f_b)^{\mu\nu} = \frac{4\pi}{c} j^\nu, \quad \partial_\mu (f_r)^{\mu\nu} = 0. \tag{49}$$

Then

$$\left(T_{(l)\,EEM}^{r}\right)^{\mu\nu} = \frac{1}{4\pi}\left[-\left(f_{(l)b} + f_{(l)r}\right)^{\mu\gamma}\left(f_{(l)b} + f_{(l)r}\right)_{\gamma}^{\nu} + \frac{1}{4} g^{\mu\nu}\left(f_{(l)b} + f_{(l)r}\right)_{\gamma\alpha}\left(f_{(l)b} + f_{(l)r}\right)^{\gamma\alpha}\right] - \frac{1}{c}\left(A_{(l)b} + A_{(l)r}\right)^{\mu} j_{(l)}^{\nu}$$

(50).

When radiation is negligible, the tensor (50) identically coincides with the EEM tensor (26), defined above.

In the next section we analyze some important physical consequences, resulting from the application of the tensor (48) and its particular form (43) to radiating and non-radiating charged particles. But now we would like to pay attention to a principal implication of Eqs. (48), (43). It is known that the Einstein equation establishes proportionality between the scalar curvature of space $R$ and the trace of the energy-momentum tensor $T$ [12]:

$$R = -\frac{8\pi\gamma_0}{c^4} T,$$

where $\gamma_0$ is the gravitational constant. One can see that the traces of tensors (43) and (48) are not equal to zero, even if the matter tensor is excluded. Hence we conclude that EM fields influence the scalar curvature of space-time via the EM masses.

## 3. Classical electrodynamics after "gauge renormalization": basic points

Below we will consider a motional equation derived from the equality

$$\partial_\mu T^{\mu\nu} = 0, \tag{51}$$

as well as the energy balance equation $\partial_\mu T^{\mu 0} = 0$ and momentum of EM field $T^{\mu 0}$, when the gauge normalized total energy-momentum tensors (43), (48) are applied.

### 3.1. Motional equation for a non-radiating charged particle

In this case we insert the tensor (43) into the conservation law (51). Then we obtain

$$\partial_\mu T^{\mu\nu} = \sum_k \left( c\partial_\mu \left[ \left( m_{(k)} + m_{(k)EM} \right) \frac{dx_{(k)}^\mu}{dt} \right] v_l^\nu + c\left( m_{(k)} + m_{(k)EM} \right) \frac{dv_{k\nu}}{dt} + \left( \partial_\mu T_{EM(ex)}^{\mu\nu} \right) \right) = 0. \tag{52}$$

The latter equation is implemented, if and only if

$$\left( m_{(k)} + m_{(k)EM} \right) \frac{dv_{l\nu}}{dt} = \frac{1}{c^2} (F_{ex})_{\nu\gamma} j_k^\gamma, \text{ and} \tag{53}$$

$$\partial_\mu \left( \left( m_{(k)} + m_{(k)EM} \right) \frac{dx_k^\mu}{dt} \right) = 0 \tag{54}$$

for each $k$.

Now consider the motion of a single non-radiating charged particle $q$ with the mechanical rest mass $M$ in an external EM field. Proceeding from continuous to discrete distributions of masses and charges, we obtain from Eq. (53)

$$(M + M_{EM}) \frac{dv_\nu}{dt} = \frac{q}{c^2} (F_{\nu\gamma})_{ex} v^\gamma, \tag{55}$$

Eq. (55) has two essential differences from the conventional motional equation (12). First it shows that a particle experiences the forces only due to the external EM fields, and a self-action is impossible. This result reflects our original exclusion of self-action from the electromagnetic energy-momentum tensor under the "gauge renormalization". Secondly, the EM mass of the particle is explicitly added to its mechanical mass. Of course, the idea to include the EM mass in the total mass of charged particles is as old as the classical model of the electron. However, it seems that this idea was usually forgotten, when the electromagnetic energy-momentum tensor and the motional equation were derived. The continuity equation (54) is common for both masses, and hence it is impossible to determine the relative contribution of $M$ and $M_{EM}$ to the total mass within classical electrodynamics. We emphasize that $M_t$ is defined in the rest frame of the particle. Eq. (55) shows that $M$, $M_{EM}$ and $M_t = M + M_{EM}$ have the identical relativistic dependence on the velocity of the particle.

Thus, we see that the introduction of a compensating term (21) into the electromagnetic energy-momentum tensor and further "gauge renormalization" remove any terms of self-action from the motional equation (55) without any changes in the Lorentz force law. In the next section we will show that for a radiating charged particle the compensating tensor (21) gives an extra-term in the Lorentz force law, which excludes any "self-acceleration" of radiating particles.

*3.2. Motional equation for a radiating charged particle*

When a particle radiates, we have to use the tensor (48) to get its motional equation. Let us assume that the external EM field, where the particle is accelerated, is described by the tensor $h^{\mu\nu}$. Then combining Eqs. (48)-(51), taking account of Eq. (24), we obtain for this particle:

$$\partial_\mu T^{\mu\nu} = \partial_\mu\left(cm\frac{dx^\mu}{dt}\frac{dx^\nu}{d\tau}\right) - h^\nu{}_\gamma j^\gamma +$$

$$\partial_\mu\left(\frac{1}{4\pi}\left[-(f_b+f_r)^{\mu\gamma}(f_b+f_r)^\nu{}_\gamma + \frac{1}{4}g^{\mu\nu}(f_b+f_r)_{\gamma\alpha}(f_b+f_r)^{\gamma\alpha}\right] - \frac{1}{c}\partial_\mu\left[(A_b+A_r)^\nu j^\mu\right]\right) =$$

$$\partial_\mu\left(cm\frac{dx^\mu}{dt}\frac{dx^\nu}{d\tau}\right) - h^\nu{}_\gamma j^\gamma + \frac{1}{4\pi}\partial_\mu\left[-f_b{}^{\mu\gamma}f_b{}^\nu{}_\gamma + \frac{1}{4}g^{\mu\nu}f_{b\gamma\alpha}f_b{}^{\gamma\alpha}\right] - \frac{1}{c}\partial_\mu(A_b{}^\nu j^\mu)$$

$$+\partial_\mu\left(\frac{1}{4\pi}\left[-f_b{}^{\mu\gamma}f_r{}^\nu{}_\gamma - f_r{}^{\mu\gamma}f_b{}^\nu{}_\gamma - f_r{}^{\mu\gamma}f_r{}^\nu{}_\gamma + \frac{1}{4}g^{\mu\nu}(f_{b\gamma\alpha}f_r{}^{\gamma\alpha} + f_{r\gamma\alpha}f_b{}^{\gamma\alpha} + f_{r\gamma\alpha}f_r{}^{\gamma\alpha})\right] - \frac{1}{c}\partial_\mu(A_r{}^\nu j^\mu)\right).$$

Noting that the EEM tensor of a bound EM field of charged particle $\frac{1}{4\pi}\left[-f_b{}^{\mu\gamma}f_b{}^\nu{}_\gamma + \frac{1}{4}g^{\mu\nu}f_{b\gamma\alpha}f_b{}^{\gamma\alpha}\right] - \frac{1}{c}(A_b{}^\mu j^\nu)$ can be replaced by the tensor of EM mass (42), we further derive:

$$\partial_\mu T^{\mu\nu} = \partial_\mu\left(c(m+m_{EM})\frac{dx^\mu}{dt}\frac{dx^\nu}{d\tau}\right) - h^\nu{}_\gamma j^\gamma - f_r{}^\nu{}_\gamma j^\gamma - \frac{1}{c}\partial_\mu(A_r{}^\nu j^\mu) = 0.$$

Taking into account the continuity equation $\partial_\mu j^\mu = 0$, and the equality $(\partial_\mu A_r{}^\nu)j^\mu = \rho dA_r{}^\nu/dt$, we obtain the motional equation in the form

$$c(m+m_{EM})\frac{dv^\nu}{dt} = h^\nu{}_\gamma j^\gamma + f_r{}^\nu{}_\gamma j^\gamma + \frac{\rho}{c}\frac{dA_r{}^\nu}{dt}. \qquad (56)$$

The first term in *rhs* of this equation describes the action of the external EM field on the particle, whereas the sum of the second and third terms represents the self-reaction of the EM radiation on the particle. The spatial component of this equation after the integration over the whole 3-space is:

$$(M+M_{EM})\frac{dv^\nu}{dt} = q\vec{E}_{ex} + q\frac{\vec{v}\times\vec{B}_{ex}}{c} + q\vec{E}_r + q\frac{\vec{v}\times\vec{B}_r}{c} + \frac{q}{c}\frac{d\vec{A}_r}{dt}. \qquad (57)$$

The last three terms in *rhs* of Eq. (57) describe the force of radiation reaction:

$$\vec{F}_r = q\vec{E}_r + q\frac{\vec{v}\times\vec{B}_r}{c} + \frac{q}{c}\frac{d\vec{A}_r}{dt}. \qquad (58)$$

This reactive force represents the sum of conventional Lorentz force $q\vec{E}_r + q\frac{\vec{v}\times\vec{B}_r}{c}$, acting on the particle due to its own EM radiation, plus the extra-term $\frac{q}{c}\frac{d\vec{A}_r}{dt}$, resulting from the radiation component of the compensating tensor (21) $\frac{1}{c}(A_r{}^\nu j^\mu)$. The presence of this extra-term essentially influences the effect of radiation reaction. Assuming the Lorenz gauge, $\nabla\vec{A} + \frac{1}{c}\frac{\partial\varphi}{\partial t} = 0$, and using the equalities $\vec{A} = -\nabla\varphi - \frac{1}{c}\frac{\partial\vec{A}}{\partial t}$, $\vec{B} = \nabla\times\vec{A}$, we can express $\vec{F}_r$ via the vector $\vec{A}$ and scalar $\varphi$ potentials as follows:

$$\vec{F}_r = q\left(-\nabla \varphi_r - \frac{\partial \vec{A}_r}{\partial t}\right) + q\frac{\vec{v} \times (\nabla \times \vec{A}_r)}{c} + \frac{q}{c}\frac{d\vec{A}_r}{dt}.$$

Using the vector identity $\vec{v} \times (\nabla \times \vec{A}) = \nabla(\vec{v} \cdot \vec{A}) - (\vec{v}\nabla)\vec{A}$ (where $\vec{v}$ is not the function of $\vec{r}$), and taking into account that $\frac{d\vec{A}_r}{dt} = \frac{\partial \vec{A}_r}{\partial t} + (\vec{v} \cdot \nabla)\vec{A}_r$, the latter equation can be transformed as

$$\vec{F}_r = -q\nabla\left(\varphi_r - \frac{(\vec{v} \cdot \vec{A}_r)}{c}\right) = -\frac{q}{\gamma}(\nabla \varphi'_r), \tag{59}$$

where $\varphi'_r$ is the scalar potential of EM radiation in the rest frame of particle, and $\gamma = 1/\sqrt{1 - v^2/c^2}$.

The obtained Eq. (59) determines the instantaneous radiation reaction force for an arbitrarily moving particle. Note that $\nabla \varphi_r$ and $\varphi_r$ have the same sign, because the electric field of EM radiation falls as $1/r$. In the non-relativistic limit we put $\gamma=1$ and $\varphi_r = \varphi'_r$. Whence, in this limit

$$\vec{F}_r = -q(\nabla \varphi_r).$$

Taking $\varphi_r = \Phi_r(r - ct)$, we obtain

$$\vec{F}_r = -q\dot{\Phi}_r \frac{\vec{r}}{r},$$

where $\dot{\Phi}_r$ is the derivative of $\Phi_r$, and $\vec{r}/r$ is the direction of observation. This equation shows that the instantaneous force, acting on a non-relativistic particle due to its EM radiation in the direction $\vec{r}/r$, is collinear with this direction.

Considering a relativistic particle, we direct the axis $x$ (ort $\vec{i}$) along its velocity $\vec{v}$. Taking into account the Lorentz transformation $x' = \gamma(x - vt)$, $y' = y$, $z = z$ (the primed coordinates refer to the rest frame of particle), we get from Eq. (59):

$$\vec{F}_r = -q\dot{\Phi}'_r \vec{i} - \frac{q\dot{\Phi}'_r}{\gamma}\vec{j} - \frac{q\dot{\Phi}'_r}{\gamma}\vec{k}. \tag{60}$$

This equation shows that the component of $\vec{F}_r$ parallel to $\vec{v}$ remains the same in the rest frame of the particle and in the laboratory frame. The components of this force orthogonal to $\vec{v}$ are reduced by the factor $1/\gamma$. Such behaviour of the force of radiation reaction completely agrees with the relativistic law of force transformation [12]. The total instantaneous force of radiation reaction is found by integration over the angular distribution of radiation. Designating this average force $\left(\bar{\vec{F}}_r\right)_\vartheta$, we notice that its resultant direction depends on the direction of the vectors $\vec{v}$ and $\dot{\vec{v}}$.

One can show that the time-like component of Eq. (56) after integration over the whole 3-space acquires the form

$$\frac{d}{dt}\frac{(M + M_{EM})}{\sqrt{1 - v^2/c^2}} = q\vec{v} \cdot \vec{E}_{ex} + \vec{v} \cdot \vec{F}_r. \tag{61}$$

Combining Eqs. (60) and (61), we arrive at

$$\frac{d}{dt}\frac{(M + M_{EM})}{\sqrt{1 - v^2/c^2}} = q\vec{v} \cdot \vec{E}_{ex} - qv\dot{\Phi}'_r. \tag{62}$$

Thus, the work done due to the force of radiation reaction properly changes the kinetic energy of the radiating particle.

At the same time, we have to emphasize that Eq. (60) does not determine the total average force, acting on a radiating charged particle, and Eq. (62) does not determine the total average work done by the radiation reaction. In order to find this total average force explicitly, we have to derive the explicit dependence of $\varphi'_r$ on the acceleration of the charged particle and to determine the angular distribution of the EM radiation. The analysis of these problems falls outside the scope of the present paper. Now we mention only that the obtained force of radiation reaction as a function of $(r - ct)$ (see, Eq. (60)) has a negative sign. It excludes any self-acceleration of the charged particle.

### 3.3. Energy flux in free and bound electromagnetic fields

First consider a free EM field in the absence of charged particles. Then the electromagnetic energy-momentum tensor (48) takes its usual form (5), and the equality $\partial_\mu T^{\mu 0} = 0$ yields:

$$\frac{\partial u}{\partial t} + \nabla \vec{S} = 0,$$

where the Poynting vector $\vec{S}$ is given by Eq. (16). If the EM radiation falls on a system of charged particles, then the latter equation transforms to Eq. (14). This equation has a reasonable physical explanation: the direction of $\vec{S}$ coincides with that of EM wave propagation, and the term $\vec{j} \cdot \vec{E}$ corresponds to absorption of EM radiation by charged particles. The same Eq. (14) is also customarily applied to a non-radiating EM field, where it leads to the appearance of a term of self-action, Eq. (14).

Now let us determine the energy balance equation for a bound EM field with the total energy-momentum tensor (43). The equality $\partial_\mu T^{\mu 0} = 0$ yields:

$$\left(\vec{j}\vec{E}\right)_{ex} + \frac{\partial u_{ex}}{c \partial t} + \nabla \cdot S_{ex} = 0, \qquad (63)$$

where $\left(\vec{j}\vec{E}\right)_{ex} = \left(F_{0\gamma}\right)_{ex} j^\gamma$ is the time rate of work done (without the self-forces), $u_{ex} = \frac{1}{4\pi}\left(-F^{0\gamma}F^0_\gamma + \frac{1}{4}F_{\gamma\alpha}F^{\gamma\alpha}\right)_{ex}$ is the part of energy density of EM field, where the "self-action" components $\vec{E}_l \vec{E}_l$ and $\vec{B}_l \vec{B}_l$ are excluded, and $\vec{S}_{ex}$ is the portion of Poynting vector, where the "self-action" components $\vec{E}_l \times \vec{B}_l$ are also excluded. It is given by the equation

$$S^i_{ex} = \frac{c}{4\pi}\left(-F^{i\gamma}F^0_\gamma\right)_{ex}.$$

In order to analyze Eq. (63) we first consider for clarity the system of two charged non-radiating particles $q_1$ and $q_2$, and subsequently generalize the results obtained to an arbitrary number $N$ of particles. For this system Eq. (63) gives:

$$\vec{j}_1\vec{E}_2 + \vec{j}_2\vec{E}_1 + \frac{\partial}{\partial t}\left(\frac{2(\vec{E}_1 \cdot \vec{E}_2) + 2(\vec{B}_1 \cdot \vec{B}_2)}{8\pi}\right) + \frac{c}{4\pi}\nabla \cdot \left[\left(\vec{E}_1 \times \vec{B}_2\right) + \left(\vec{E}_2 \times \vec{B}_1\right)\right] = 0. \qquad (64)$$

Let us transform the last term in *lhs* of Eq. (64), applying the vector identity $\nabla \cdot (\vec{a} \times \vec{b}) = \vec{b} \cdot (\nabla \times \vec{a}) - \vec{a} \cdot (\nabla \times \vec{b})$. Then we obtain
$$\nabla \cdot \left(\vec{E}_1 \times \vec{B}_2 + \vec{E}_2 \times \vec{B}_1\right) = \vec{B}_2 \cdot (\nabla \times \vec{E}_1) - \vec{E}_1 \cdot (\nabla \times \vec{B}_2) + \vec{B}_1 \cdot (\nabla \times \vec{E}_2) - \vec{E}_2 \cdot (\nabla \times \vec{B}_1).$$
The relationship between electric and magnetic fields in the bound EM field is
$$\vec{B} = \vec{v} \times \vec{E}/c.$$

Combining the last two equations and using the Maxwell equation $\nabla \times \vec{E} = -\frac{1}{c}\frac{\partial \vec{B}}{\partial t}$ as well, we get

$$c\nabla \cdot \left(\vec{E}_1 \times \vec{B}_2 + \vec{E}_2 \times \vec{B}_1\right) = \left[-\vec{B}_2 \cdot \frac{\partial \vec{B}_1}{\partial t} - \vec{E}_1 \cdot \left(\nabla \times (\vec{v}_2 \times \vec{E}_2)\right) - \vec{B}_1 \cdot \frac{\partial \vec{B}_2}{\partial t} - \vec{E}_2 \cdot \left(\nabla \times (\vec{v}_1 \times \vec{E}_1)\right)\right].$$

Applying the vector identity $\nabla \times (\vec{v} \times \vec{E}) = \vec{v} \cdot (\nabla \cdot \vec{E}) - (\vec{v} \cdot \nabla)\vec{E}$, the Maxwell equations $\vec{E} = 4\pi\rho$, $\frac{1}{c}\frac{\partial \vec{E}}{\partial t} = \nabla \times \vec{B} - \frac{4\pi}{c}\vec{j}$, and taking into account that $\vec{j} = \rho\vec{v}$, we further derive

$$c\nabla \cdot (\vec{E}_1 \times \vec{B}_2 + \vec{E}_2 \times \vec{B}_1) = -\vec{B}_2 \cdot \left(\vec{v}_1 \times \frac{\partial \vec{E}_1}{c\partial t}\right) - \vec{E}_1 \cdot [\vec{v}_2 \cdot (\nabla \vec{E}_2) - (\vec{v}_2 \cdot \nabla)\vec{E}_2] -$$

$$-\vec{B}_1 \cdot \left(\vec{v}_2 \times \frac{\partial \vec{E}_2}{c\partial t}\right) - \vec{E}_2 \cdot [\vec{v}_1 \cdot (\nabla \vec{E}_1) - (\vec{v}_1 \cdot \nabla)\vec{E}_1] =$$

$$= -\vec{B}_2 \cdot \left[\vec{v}_1 \times \left(\nabla \times \vec{B}_1 - \frac{4\pi}{c}\vec{j}_1\right)\right] - 4\pi\vec{E}_1 \cdot \vec{j}_2 + \vec{E}_1 \cdot [(\vec{v}_2 \cdot \nabla)\vec{E}_2] -$$

$$-\vec{B}_1 \cdot \left[\vec{v}_2 \times \left(\nabla \times \vec{B}_2 - \frac{4\pi}{c}\vec{j}_2\right)\right] - 4\pi\vec{E}_2 \cdot \vec{j}_1 + \vec{E}_2 \cdot [(\vec{v}_1 \cdot \nabla)\vec{E}_1] =$$

$$= \vec{B}_2 \cdot [(\vec{v}_1 \cdot \nabla)\vec{B}_1] - 4\pi\vec{E}_1 \cdot \vec{j}_2 + \vec{E}_1 \cdot [(\vec{v}_2 \cdot \nabla)\vec{E}_2] + \vec{B}_1 \cdot [(\vec{v}_2 \cdot \nabla)\vec{B}_2] - 4\pi\vec{E}_2 \cdot \vec{j}_1 + \vec{E}_2 \cdot [(\vec{v}_1 \cdot \nabla)\vec{E}_1]. \quad (65)$$

Under manipulations with the latter equation we take into account that $\vec{v}_1 \times \vec{j}_1 = \vec{v}_2 \times \vec{j}_2 = 0$, and $\vec{v}_1 \cdot \vec{B}_1 = \vec{v}_2 \cdot \vec{B}_2 = 0$. Combining Eqs. (65) and (64), we arrive at the equality

$$\frac{\partial}{\partial t}\frac{2(\vec{E}_1 \cdot \vec{E}_2) + 2(\vec{B}_1 \cdot \vec{B}_2)}{8\pi} + \frac{\vec{E}_1 \cdot [(\vec{v}_2 \cdot \nabla)\vec{E}_2] + \vec{B}_1 \cdot [(\vec{v}_2 \cdot \nabla)\vec{B}_2] + \vec{E}_2 \cdot [(\vec{v}_1 \cdot \nabla)\vec{E}_1] + \vec{B}_2 \cdot [(\vec{v}_1 \cdot \nabla)\vec{B}_1]}{4\pi} = 0.$$
(66)

The obtained Eq. (66) does not yet determine the total flow of energy in a bound EM field, because the flow of EM masses should be added. As we mentioned above, due to the fixed ratio of mechanical to EM mass (the law of charge conservation), the continuity equation (54) is separately valid for the density of EM mass $u_s/c^2$:

$$\partial_\mu \left(u_s \frac{dx_s^\mu}{dt}\right) = 0.$$

For the considered case of two charged particles we get for the densities of their EM masses:

$$\frac{\partial}{\partial t}\left(\frac{E_1^2 + B_1^2 + E_2^2 + B_2^2}{8\pi}\right) + \nabla\left[\frac{\vec{v}_1(E_1^2 + B_1^2) + \vec{v}_2(E_2^2 + B_2^2)}{8\pi}\right] = 0. \quad (67)$$

The total flow of EM energy is determined by summing up of Eqs. (66) and (67):

$$\frac{\partial u_\Sigma}{\partial t} + \nabla\left[\frac{\vec{v}_1(E_1^2 + B_1^2) + \vec{v}_2(E_2^2 + B_2^2)}{8\pi}\right] +$$
$$+ \frac{\vec{E}_1 \cdot [(\vec{v}_2 \cdot \nabla)\vec{E}_2] + \vec{B}_1 \cdot [(\vec{v}_2 \cdot \nabla)\vec{B}_2] + \vec{E}_2 \cdot [(\vec{v}_1 \cdot \nabla)\vec{E}_1] + \vec{B}_2 \cdot [(\vec{v}_1 \cdot \nabla)\vec{B}_1]}{4\pi} = 0, \quad (68)$$

where $u_\Sigma = \frac{E_\Sigma^2 + B_\Sigma^2}{8\pi}$ is the energy density of the total EM field of two particles $q_1$ and $q_2$. Here we denote the resultant electric and magnetic fields as $\vec{E}_\Sigma = \vec{E}_1 + \vec{E}_2$, $\vec{B}_\Sigma = \vec{B}_1 + \vec{B}_2$. Further, taking into account the equality $\nabla \cdot [\vec{v}(E^2 + B^2)] = 2\vec{E} \cdot [(\vec{v} \cdot \nabla)\vec{E}] + 2\vec{B} \cdot [(\vec{v} \cdot \nabla)\vec{B}]$, we can join the last two terms in *lhs* of Eq. (68) into a single one:

$$\vec{E}_1 \cdot [(\vec{v}_1 \cdot \nabla)\vec{E}_1] + \vec{B}_1 \cdot [(\vec{v}_1 \cdot \nabla)\vec{B}_1] + \vec{E}_2 \cdot [(\vec{v}_2 \cdot \nabla)\vec{E}_2] + \vec{B}_2 \cdot [(\vec{v}_2 \cdot \nabla)\vec{B}_2] +$$
$$+ \vec{E}_1 \cdot [(\vec{v}_2 \cdot \nabla)\vec{E}_2] + \vec{B}_1 \cdot [(\vec{v}_2 \cdot \nabla)\vec{B}_2] + \vec{E}_2 \cdot [(\vec{v}_1 \cdot \nabla)\vec{E}_1] + \vec{B}_2 \cdot [(\vec{v}_1 \cdot \nabla)\vec{B}_1] =$$
$$\vec{E}_\Sigma \cdot [(\vec{v}_1 \cdot \nabla)\vec{E}_1] + \vec{B}_\Sigma \cdot [(\vec{v}_1 \cdot \nabla)\vec{B}_1] + \vec{E}_\Sigma \cdot [(\vec{v}_2 \cdot \nabla)\vec{E}_2] + \vec{B}_\Sigma \cdot [(\vec{v}_2 \cdot \nabla)\vec{B}_2].$$

Then Eq. (68) acquires the form

$$\frac{\partial u_\Sigma}{\partial t} + \frac{\vec{E}_\Sigma \cdot [(\vec{v}_1 \cdot \nabla)\vec{E}_1] + \vec{B}_\Sigma \cdot [(\vec{v}_1 \cdot \nabla)\vec{B}_1] + \vec{E}_\Sigma \cdot [(\vec{v}_2 \cdot \nabla)\vec{E}_2] + \vec{B}_\Sigma \cdot [(\vec{v}_2 \cdot \nabla)\vec{B}_2]}{4\pi} = 0.$$

Introducing the partial spatial operator $\nabla_{-\Sigma}$, acting only on the electric and magnetic fields of the first and second particle, but not on the total EM fields, the latter equation can be written in a compact form

$$\frac{\partial u_\Sigma}{\partial t} + \frac{\nabla_{-\Sigma} \cdot [\vec{v}_1(\vec{E}_\Sigma \cdot \vec{E}_1 + \vec{B}_\Sigma \cdot \vec{B}_1) + \vec{v}_2(\vec{E}_\Sigma \cdot \vec{E}_2 + \vec{B}_\Sigma \cdot \vec{B}_2)]}{4\pi} = 0, \tag{69}$$

Now consider the case of an arbitrary number $N$ of charged particles. One can show that for this case Eqs. (66) and (67) transform to

$$\frac{\partial}{\partial t}\left[\frac{1}{4\pi}\sum_{k \neq k'}(\vec{E}_k \cdot \vec{E}_{k'}) + (\vec{B}_k \cdot \vec{B}_{k'})\right] + \frac{1}{4\pi}\sum_{k \neq k'}\{\vec{E}_k \cdot [(\vec{v}_{k'} \cdot \nabla)\vec{E}_{k'}] + \vec{B}_k \cdot [(\vec{v}_{k'} \cdot \nabla)\vec{B}_{k'}]\} = 0, \tag{70}$$

$$\frac{\partial}{\partial t}\left[\frac{1}{8\pi}\sum_k (E_k^2 + B_k^2)\right] + \nabla\left[\frac{1}{8\pi}\sum_k \vec{v}_k (E_k^2 + B_k^2)\right] = 0. \tag{71}$$

Eq. (71) describes the flow of EM energy of the particles associated with their EM masses, whereas Eq. (70) can be interpreted as the flow of energy resulting from the vector addition of bound EM fields of different particles. The sum of Eqs. (70) and (71) gives the total flow of energy in a bound EM field:

$$\frac{\partial u_\Sigma}{\partial t} + \frac{1}{4\pi}\nabla_{-\Sigma} \cdot \sum_{k=1}^{N} \vec{v}_k (\vec{E}_\Sigma \cdot \vec{E}_k + \vec{B}_\Sigma \cdot \vec{B}_k) = 0, \text{ or}$$

$$\frac{\partial u_\Sigma}{\partial t} + \nabla_{-\Sigma} \cdot \vec{U}_G = 0, \tag{72}$$

where we have introduced the vector

$$\vec{U}_G = \sum_{k=1}^{N} \vec{v}_k (\vec{E}_\Sigma \cdot \vec{E}_k + \vec{B}_\Sigma \cdot \vec{B}_k), \tag{73}$$

and $\vec{E}_\Sigma = \sum_k \vec{E}_k$ and $\vec{B}_\Sigma = \sum_k \vec{B}_k$ are the resultant electric and magnetic fields.

Thus, we have got the energy balance equation (72), which determines the energy flux in a bound EM field. First of all, we see that it does not contain the term of dissipation of EM energy $\vec{j} \cdot \vec{E}$. In this connection we mention that the term $\vec{j} \cdot \vec{E}$ describes a time derivative of the kinetic energy of particles, which is equal with the opposite sign to the time rate of change of potential energy of particles in the bound EM field. In turn, the change of potential energy is already included in the partial time derivative $\partial u/\partial t$. Hence, in comparison with the energy balance equation (14) for free EM field, the term $\vec{j} \cdot \vec{E}$ does not appear for the bound fields. Inasmuch as Eq. (72) represents the sum of Eqs. (70), (71), it incorporates two different effects: the flow of EM masses of all individual particles, as well as the superposition of bound EM fields of the particles.

Eq. (72) was first obtained in ref. [13] as a formal solution of Maxwell's equations. However, the physical meaning of this equation was not clarified. We notice that in the particular case, where the instantaneous velocities of all particles are equal to each other ($\vec{v}_k = \vec{v}$ for any $k$), Eq. (72) acquires the form

$$\frac{\partial u_\Sigma}{\partial t} + \nabla \cdot (\vec{v} u_\Sigma) = 0. \tag{74}$$

This equation shows that the resultant EM field rigidly moves together with the source particles. It seems interesting that each individual particle carries its EM mass independently of other particles, but the superposition of bound EM fields from all particles transforms the sum of these individual motions into a common motion of the resultant bound EM field at the same velocity $\vec{v}$. The vector $\vec{U} = \vec{v} u_\Sigma$ was first introduced by Umov in fluid mechanics more than one century ago [14]. Hence we can name the vector (73) as the generalized Umov's vector, which describes the energy fluxes in the system of charged particles, moving at different velocities.

Thus, the energy balance equation resulting from the total energy-momentum tensor (43) for a bound EM field completely differs from the Poynting expression both in form and physical interpretation. The scientific literature on classical EM theory contains numerous problems where the energy fluxes in non-radiative EM fields, guided by the Poynting vector, give strange physical pictures. But what is more important, there are problems, hitherto ignored, where the Poynting vector totally fails to describe the energy flux in a bound EM field. For example, consider the motion of a charged parallel plate capacitor in the direction normal to the plates (Fig. 1). The square plates have the size $a \times a$, where $a >> d$, $d$ being the distance between the plates. Then in the inner space region far from the boundaries of the plates, the electric field $\vec{E}$ is constant and coincides with the direction of velocity of the plates $\vec{v}$. Since the magnetic field is absent between the plates ($\vec{v} \times \vec{E} = 0$), the generalized Umov's vector is equal to $\vec{U}_{UG} = \vec{v} E^2/8\pi$: the electric field rigidly moves together with the plates. However, in no way can this result be understood with the Poynting vector. Indeed, in this space region $\vec{S} = c(\vec{E} \times \vec{B})/4\pi = 0$, and there is no energy flux inside the capacitor in the Poynting sense. Thus, only Eq. (72) adequately describes the energy flux inside the capacitor, where the EM field rigidly moves with the plates.

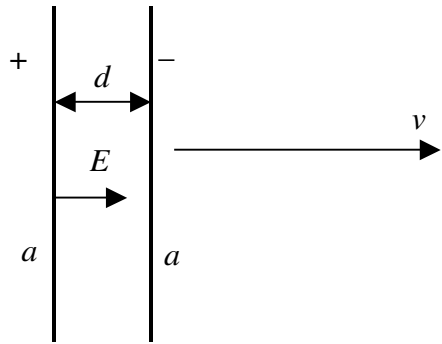

Fig. 1. A parallel plate charged capacitor moves at constant velocity $\vec{v}$ along the normal to the plates.

Let us show that this result complies with the finite (light) velocity of propagation of the EM field. It is known that the expression for an electric and magnetic field of a moving particle contains two parts: the velocity-dependent term and acceleration-dependent term. The velocity-dependent term can be written in "present time coordinates", which yield the Heaviside ellipsoid [6-8]. In such "present time coordinates" the bound EM field rigidly propagates together with the moving source particle. We see that the generalized Umov's vector gives the same result, when applied to a single charged particle.

The results obtained in this sub-section indicate that free and bound EM fields have substantially different physical properties. It warrants their primary distinction in the original energy-momentum tensor (48).

*3.4. The momentum of free and bound EM fields*

The momentum density of the EM field is the component $T_{EM}^{i0}/c$ ($i=1\ldots3$) in the electromagnetic energy-momentum tensor. For electromagnetic radiation it is written in the known form

$$\vec{p} = \frac{\vec{E} \times \vec{B}}{4\pi c}. \tag{75}$$

For a bound EM field we determine the electromagnetic energy-momentum tensor as

$$T_{EM}{}^{\mu\nu} = \sum_k m_{(k)EM} \frac{d\,x^{\mu}_{(k)}}{dt}\frac{d\,x^{\nu}_{(k)}}{d\tau} + \frac{1}{4\pi}\left(-F^{\mu\gamma}F^{\nu}_{\gamma} + \frac{1}{4}g^{\mu\nu}F_{\gamma\alpha}F^{\gamma\alpha}\right)_{ex}.$$

which is derived from the tensor Eq. (43) by the exclusion of its mechanical part. Then the momentum density as a function of velocities of particles and their EM fields is

$$\vec{p} = \sum_{k=1}^{N} \frac{\vec{v}_k (E_k^2 + B_k^2)}{c^2\sqrt{1 - v_k^2/c^2}} + \sum_{k \neq k'} \frac{\vec{E}_k \times \vec{B}_{k'}}{4\pi}. \tag{76}$$

The total momentum of a bound EM field is computed by integration of Eq. (76) over the whole 3-space:

$$\vec{P}_{EM} = \int_V \frac{\sum_k \vec{v}_k (E_k^2 + B_k^2)}{c^2\sqrt{1 - v_k^2/c^2}} dV + \int_V \sum_{k \neq k'} \frac{\vec{E}_k \times \vec{B}_{k'}}{4\pi} dV. \tag{77}$$

It consists of two parts: the momentum density, associated with the EM mass of charged particles, and the momentum density, resulting from the superposition of EM fields of different particles. Before considering the forces and mutual transformation of mechanical and EM momentum in systems of charged particles, we emphasize that the first term in *rhs* of Eq. (76) represents a contribution of EM momentum of the particle, associated with its EM mass, to the total momentum of that particle. Therefore, the time rate of the first term in *rhs* of Eq. (77) is rather the consequence than the cause of the force experienced by the particle. Hence the external forces, acting on charged particles, are determined by the time rate of the second term in *rhs* of Eq. (77).

Let us consider an isolated system, consisting of two non-radiating charged particles $q_1$ and $q_2$, and determine a total force exerted on this system. In general, it does not vanish, owing to violation of Newton's third law in EM interactions. Adding the mechanical momenta of both particles to Eq. (77), we obtain

$$\vec{P}_{EM} = \frac{(M_1 + M_{EM1})\vec{v}_1}{\sqrt{1 - v_1^2/c^2}} + \frac{(M_2 + M_{EM2})\vec{v}_2}{\sqrt{1 - v_1^2/c^2}} + \int_V (\vec{E}_1 \times \vec{B}_2 + \vec{E}_2 \times \vec{B}_1) dV.$$

The resulting force, acting on the particles, is

$$\vec{F} = \frac{d}{dt}\left[\frac{(M_1 + M_{EM1})\vec{v}_1}{\sqrt{1 - v_1^2/c^2}} + \frac{(M_2 + M_{EM2})\vec{v}_2}{\sqrt{1 - v_1^2/c^2}}\right] = -\frac{d}{dt}\int_V (\vec{E}_1 \times \vec{B}_2 + \vec{E}_2 \times \vec{B}_1) dV. \tag{78}$$

If the particles are non-relativistic, then [15]

$$\int_V (\vec{E}_1 \times \vec{B}_2) dV = \frac{q_1 \vec{A}_{21}}{c}, \quad \int_V (\vec{E}_2 \times \vec{B}_1) dV = \frac{q_2 \vec{A}_{12}}{c},$$

where $\vec{A}_{21}$ is the vector potential produced by the particle 2 at the location of particle 1, and $\vec{A}_{12}$ is the vector potential of particle 1 at the location of particle 2. Hence

$$\vec{F} = \frac{d\vec{P}_1}{dt} + \frac{d\vec{P}_2}{dt} = -\frac{q_1}{c}\frac{d\vec{A}_{21}}{dt} - \frac{q_2}{c}\frac{d\vec{A}_{12}}{dt}. \tag{79}$$

This equation reflects the law of conservation of the canonical momentum

$$\vec{P}_C = \left(\vec{P}_1 + \frac{q_1 \vec{A}_{21}}{c}\right) + \left(\vec{P}_2 + \frac{q_2 \vec{A}_{12}}{c}\right) = \text{const}$$

for the considered non-radiating non-relativistic system. Eq. (79) has also been derived in [13] within the Lagrangian formalism.

Without the "gauge renormalization", the conventional Poynting vector would determine the resultant force:

$$\vec{F} = \frac{d}{dt}\left(\frac{M_1\vec{v}_1}{\sqrt{1-v_1^2/c^2}} + \frac{M_2\vec{v}_2}{\sqrt{1-v_1^2/c^2}}\right) = -\frac{d}{dt}\int_V (\vec{E}_1\times\vec{B}_1 + \vec{E}_1\times\vec{B}_2 + \vec{E}_2\times\vec{B}_1 + \vec{E}_2\times\vec{B}_1)dV, \quad (80)$$

and instead of Eq. (79), we would obtain

$$\vec{F} = \frac{d\vec{P}_1}{dt} + \frac{d\vec{P}_2}{dt} = -q_1\frac{d\vec{A}_{21}}{dt} - q_2\frac{d\vec{A}_{12}}{dt} - \frac{d}{dt}\int_V(\vec{E}_1\times\vec{B}_1)dV - \frac{d}{dt}\int_V(\vec{E}_2\times\vec{B}_2)dV. \quad (81)$$

which does not agree with the law of conservation of the canonical momentum. Moreover, at the location of point-like charges the terms $(\vec{E}_1\times\vec{B}_1)$ and $(\vec{E}_2\times\vec{B}_2)$ increase as $1/r^4$ at $r\to 0$, and the third and fourth integrals in *rhs* of Eq. (81) diverge. The difference between Eqs. (79) and (81) reflects a physical meaning of the "gauge renormalization", when the time rates of the terms, taken from the same source particles ($(\vec{E}_1\times\vec{B}_1)$ and $(\vec{E}_2\times\vec{B}_2)$) contribute to their own EM momentum, associated with the EM mass, and thus represent the consequences of an action of the external forces, but not their cause.

Now consider stationary EM systems, where the problem of mutual transformation between mechanical and EM momentum was explored at both theoretical and experimental levels (see, e.g., [16-18]). Such systems are presented fundamentally as an electrically neutral magnetic dipole with the moment $\vec{\mu}$ and a resting charged particle $q$. The EM momentum density of this system is proportional to the vector product $\vec{E}_q\times\vec{B}_\mu$, where both fields are necessarily taken from the different sources. Hence the total EM momentum of the system is the same under the conventional Poynting's approach and within classical electrodynamics after the gauge renormalization.

Finally, for an isolated charged particle, moving at the constant velocity $\vec{v}$ in a laboratory, the momentum density of the bound EM field is determined as

$$\vec{p}_{EM} = \vec{v}\frac{u(v=0)}{c^2\sqrt{1-v^2/c^2}} = m_{EM}(\vec{v})\vec{v}, \quad (82)$$

where $m_{EM}(\vec{v}) = \dfrac{u(v=0)}{c^2\sqrt{1-v^2/c^2}}$ is the density of velocity-dependent EM mass of the particle. Since the equality $u(\vec{v}) = m_{EM}(\vec{v})c^2$ is implemented by definition, then the known problem "4/3" is formally eliminated in Eq. (82). It does not mean that the problem is resolved: it is simply relocated from Eq. (82) into Eqs. (46) and (47). It reflects the obvious fact that any gauge operation does not change the total energy of charged particles, which includes the energy that provides the stability of the electron ("Poincaré stresses" [19]). The following subsection is devoted to closer classical analysis of this problem with Eqs. (46) and (47), obtained under the "gauge renormalization".

*3.5. Proper mass of electron and "Poincaré stresses"*

In section 2 we mentioned that a finiteness of the electromagnetic mass of the electron could result as the difference of two divergent integrals in Eq. (46). Note that the second term in *rhs* of Eq. (46) has a negative sign. This fact already displays a presence of some non-electromagnetic forces, such as the "Poincaré stresses": otherwise (for positive sign of both terms in *rhs* of Eq. (46)), the stability of the electron would be impossible. Nevertheless, we continue the classical analysis of this equation and assume the conventional expressions for the scalar potential and electric field,

$$\varphi = q/r,\; E = q/r^2, \quad (83), (84)$$

correspondingly, in the spatial region where the change density of the electron $\rho$ can be taken equal to zero. At distances $r<r_b$ from the "center of electron" ($r_b$ is some boundary value) the charge density is not zero, and we cannot adopt Eqs. (83), (84). Hence in the non-relativistic limit

18$$M_E c^2 = \frac{q^2}{2r_b} + \int_{V(r<r_b)} \frac{E^2}{8\pi} dV - \int_{V(r<r_b)} \rho\varphi \, dV, \tag{85}$$

$$M_M \vec{v} = \frac{2q^2 \vec{v}}{3r_b c^2} + \int_{V(r<r_b)} \frac{1}{4\pi c}(\vec{E}\times\vec{B}) dV - \int_{V(r<r_b)} \left(\frac{\vec{v}\rho\varphi}{c^2}\right) dV. \tag{86}$$

Manipulating with the latter equations, we have used the known results [11]

$$\int_{r_b}^{\infty} \frac{E^2}{8\pi} dV = \frac{q^2}{2r_b}, \quad \int_{r_b}^{\infty} \frac{1}{4\pi}(\vec{E}\times\vec{B}) dV = \frac{2\vec{v}}{3c^2}\int_{r_b}^{\infty} E^2 r^2 dr = \frac{2q^2 \vec{v}}{3r_b c^2}.$$

In Eqs. (85) and (86) we separately designated the EM masses, determined from the energy ($M_E$) and momentum ($M_M$) conservation laws. The ratio of numerical coefficients in the first terms of *rhs* of Eq. (85) and Eq. (86) is equal to 4/3, which gives rise to the familiar problem. However, one can see that an actual ratio $M_M/M_E$ is determined by all terms in Eq. (85) and (86). Hence we have to evaluate the remaining integrals in these equations, where $\rho \neq 0$ within the volume of integration. We can certainly state that due to the isotropy of space, $\rho$ is a function of $r$ only [5]. Therefore, if the Maxwell's equations and the Gauss theorem are applicable to very small distances, then

$$E(r) = \frac{4\pi}{r^2}\int_0^r \rho(r)r^2 dr, \text{ and } \varphi(r) = 4\pi \int_0^r \frac{1}{r^2}\left[\int_0^r \rho(r)r^2 dr\right] dr.$$

Inserting these values into Eqs. (85), (86), we get

$$M_E c^2 = \frac{q^2}{2r_b} + \int_{V(r<r_b)} \frac{2\pi}{r^4}\left[\int_0^r \rho(r)r^2 dr\right]^2 dV - \int_{V(r<r_b)} \rho(r)\int_0^r \frac{4\pi}{r^2}\left[\int_0^r \rho(r)r^2 dr\right] dr\, dV, \tag{87}$$

$$M_M \vec{v} = \frac{2q^2\vec{v}}{3r_b c^2} + \frac{2\vec{v}}{3c^2}\int_{V(r<r_b)} \frac{(4\pi)^2}{r^2}\left(\int_0^r \rho(r)r^2 dr\right)^2 dr - \frac{\vec{v}}{c^2}\int_{V(r<r_b)} \rho(r)\int_0^r \frac{4\pi}{r^2}\left[\int_0^r \rho(r)r^2 dr\right] dr\, dV. \tag{88}$$

Thus, the ratio of $M_M$ to $M_E$ depends on the shape of the function $\rho = \rho(r)$. In general, in any other classical model of electron, the ratio $M_M/M_E$ is also model-dependent. However, now there is a qualitatively new property of Eqs. (87), (88), which is absent in the known classical models of the electron: namely, the existence of "zero solutions", when both $M_M$ to $M_E$ masses simultaneously go to zero under a proper choice of the function $\rho(r)$.

In order to prove an availability of such "zero solutions", we choose $\rho = const \cdot r^n$ ($n$ is some number), and use the condition of normalization $\int_{V(r<r_b)} \rho dV = q$. Then we get after straightforward calculations:

$$M_E c^2 = \frac{q^2}{2r_b} - \frac{q^2}{2r_b}\frac{(n+4)}{(n+2)(2n+5)}, \tag{89}$$

$$M_M \vec{v} = \frac{2q^2\vec{v}}{3r_b c^2} - \frac{q^2\vec{v}(n+5)}{3r_b c^2 (n+2)(2n+5)}. \tag{90}$$

We see that for $n=-3$ ($\rho = \frac{const}{r^3}$), both masses are vanished: $M_E=M_M=0$[2]. This result could mean that the total observable mass of electron is not electromagnetic in nature, and determined by its mechanical mass and the energy of the stresses of Poincaré. It resolves the problem of "4/3", because the non-electromagnetic masses and momenta are related by the conventional relativistic for-mula. We cannot even exclude that the "Poincaré stresses" are completely responsible for the observable electron's mass. In addition, both masses $M_E$, $M_M$ vanish at any $r_b$, including the limit $r_b \to 0$ ("point-like" elementary particle). Of course, the designated problems lie beyond the limit of validity of classical physics. Nevertheless, the choice $n=-3$ in Eqs. (89), (90) gives a possible classical resolution of the problem "4/3", which was originally formulated within classical physics, too.

---

[2] Although the normalization integral $\int \rho dV = q$ diverges for the function $\rho = \frac{const}{r^3}$, this defect can be ignored within classical physics.

We underline that the vanishing of the electron's EM mass results from the equality of EM energy $\int (E^2+B^2/8\pi)dV$, integrating over the whole space, and the "potential energy" $\int \rho\varphi dV$, integrating over the "inner volume" of the electron. Therefore, in the outer (with respect to the electron) space, where $\rho=0$, the mass density is equal to $u/c^2$ as before, where $u$ is given by Eq. (15). Hence there is no need to re-define the parameters of mass density $m$, $m_{EM}$ in the energy-momentum tensors (43) and (48), even if the sum $m+m_{EM}$ is exactly equal to the mass of "Poincaré stresses".

## 4. Conclusions

1. In this paper we have removed the inconsistency that existed up to now in classical electrodynamics. Namely, in the gauge transformation of canonical energy-momentum (1) to the symmetric form, we applied the non-homogeneous Maxwell equation (10) instead of the irrelevant homogeneous equation (6). As a result, the symmetric "generalized" energy-momentum tensor acquired the additional "compensating term" (21). The presence of that "compensating term" allows a gauge transformation, converting the divergent terms of classical electrodynamics (infinite self-force, self-energy and self-momentum) to converging integrals. This operation was named as "gauge renormalization".

2. The obtained energy-momentum tensor (48) with its particular form (43) for a bound EM field has been applied to the fundamental problems of classical electrodynamics: the motional equation, the energy balance equation, and the momentum conservation law for the system of moving charged particle.

The motional equation for a non-radiating charged particle does not contain its self-force, and the mass parameter represents the sum of mechanical and electromagnetic masses. The motional equation for a radiating particle also contains the sum of mechanical and electromagnetic masses, and does not yield any "runaway solutions".

The energy flux in a free EM field is guided by the Poynting vector. The energy flux in a bound EM field is described by the generalized Umov vector, defined in the paper. This result shows that free and bound EM fields have substantially different physical properties, which warrant their primary distinction in the energy-momentum tensor (48).

3. According to the "gauge renormalization", a relationship between the mass density and the momentum density of a bound EM field has the conventional relativistic form, whereas the problem of "Poincaré stresses" is relocated to the converging integrals, which determine the total EM mass and momentum of a charged particle (Eqs. (46), (47)). At the classical level, these integrals depend on the charge density distribution $\rho(r)$ in the "inner volume" of electron. It has been shown that variation of the shape of the function $\rho(r)$ allows making zero both masses $M_E$ and $M_M$ simultaneously. In this case the total mass of electron has a non-electromagnetic origin and as a limiting case, it could be fully ascribed to the energy of Poincaré stresses. For such non-electromagnetic mass the problem of "4/3" disappears.

## Acknowledgment

The author warmly thanks Oleg V. Missevitch (Belarusian State University) and Thomas E. Phipps, Jr. (Urbanna, Illinois, USA) for careful reading of the manuscript and for useful remarks, which have been taken into account in the final version of the paper.

**References**

1. H.A. Lorentz. *The theory of Electrons*. 2$^{nd}$ Ed. (Dover, 1952).
2. M. Abraham. Ann. Phys. **10** (1903) 105.
3. P.A.M. Dirac. Proc. Roy. Soc. (London) **A167** (1938) 148.
4. M. Born and L. Infeld. Proc. Roy. Soc. (London) **A144** (1934) 145.
5. F. Rohrlich. *Classical Charged Particles*, (Reading, Mass.: Addison-Wesley, 1965).
6. L.D. Landau and E.M. Lifshitz. *The Classical Theory of Fields*, 2nd edn (New York: Pergamon Press, 1962).
7. J.D. Jackson. *Classical Electrodynamics*. (New York, Wiley, 1975).
8. W.K.H. Panofsky and M. Phillips *Classical Electricity and Magnetism* 2nd edn (Reading, Mass., Addison-Wesley, 1962)
9. W. Pauli. Principles of Quantum Mechanics, Encylcopedia of Physcs, Vol. V/1 (Springer, Berlin, 1958).
10. E.J. Moniz and D.H. Sharp. Phys. Rev. **15** (1977) 2850.
11. R.P. Feynman, R.B. Leighton, and M. Sands. The Feynman Lectures in Physics. Vol. 2, Addison-Wesley, Reading, Mass. (1964).
12. C. Møller. *The Theory of Relativity*. Clarendon Press, Oxford (1972).
13. A.L. Kholmetskii. Annales de la Foundation Louis de Broglie, **29** (2004) 549.
14. N.A. Umov. Izbrannye Sochineniya (selected works), Gostechizdat, Moscow (1950) (in Russian).
15. J.M. Aguirregabiria. A. Hernández and M. Rivas. Eur. J. Phys. 3 (1982) 30-33.
16. W. Shockley and R. James. Phys. Rev. Lett., **18**, 876, (1967).
17. Y. Aharonov, P. Pearle and L. Vaidman. Phys. Rev., **A37**, 4052, (1988).
18. M.Graham and D.G.Lahoz. Nature, **285** (1980) 154.
19. H. Poincaré. Rend. Circ. Mat. Palermo **21** (1906) 129. (Engl. trans. with modern notation in H.M. Schwartz. Am. J. Phys. **40** (1972) 860).